\documentclass[journal]{IEEEtran}
\usepackage{blindtext}
\usepackage{graphicx}

\ifCLASSINFOpdf
\else
\fi

\usepackage[cmex10]{amsmath}
\usepackage{amssymb}
\usepackage{color}
\usepackage{cite}
\usepackage{arydshln} 
\newcommand\norm[1]{\left\lVert#1\right\rVert} 
\usepackage{soul}
\usepackage{tikz}
\usetikzlibrary{shapes, arrows, automata, positioning}
\tikzstyle{startstop} = [rectangle, rounded corners, minimum width=0.5cm, minimum height=0.8cm,text centered, draw=black, fill=green!10]
\tikzstyle{io} = [trapezium, trapezium left angle=70, trapezium right angle=110, minimum width=0.5cm, minimum height=0.8cm, text centered, draw=black, fill=blue!30]
\tikzstyle{process} = [rectangle, minimum width=0.5cm, minimum height=0.8cm, text centered, draw=black, fill=white!100]
\tikzstyle{decision} = [diamond, minimum width=0.5cm, minimum height=0.8cm, text centered, draw=black, fill=yellow!20]
\tikzstyle{arrow} = [thick,->,>=stealth]

\begin{document}

\title{A robust extended Kalman filter for power system dynamic state estimation using PMU measurements}

\author{Marcos~Netto$^{1}$,~\IEEEmembership{Student~Member,~IEEE,}
        Junbo~Zhao$^{1,2}$,~\IEEEmembership{Student~Member,~IEEE,}
        and~Lamine~Mili$^{1}$,~\IEEEmembership{Fellow,~IEEE}
\thanks{$^{1}$The Bradley Department of Electrical and Computer Engineering, Virginia Polytechnic Institute and State University, Northern Virginia Center, Falls Church (VA), USA (e-mail: \{mnetto, junbob, lmili\}@vt.edu).}
\thanks{$^{2}$School of Electrical Engineering, Southwest Jiaotong University, Chengdu, China.}}

\maketitle

\begin{abstract}
This paper develops a robust extended Kalman filter to estimate the rotor angles and the rotor speeds of synchronous generators of a multimachine power system. Using a batch-mode regression form, the filter processes together predicted state vector and PMU measurements to track the system dynamics faster than the standard extended Kalman filter. Our proposed filter is based on a robust GM-estimator that bounds the influence of vertical outliers and bad leverage points, which are identified by means of the projection statistics. Good statistical efficiency under the Gaussian distribution assumption of the process and the observation noise is achieved thanks to the use of the Huber cost function, which is minimized via the iteratively reweighted least squares algorithm.  The asymptotic covariance matrix of the state estimation error vector is derived via the covariance matrix of the total influence function of the GM-estimator.  Simulations carried out on the IEEE 39-bus test system reveal that our robust extended Kalman filter exhibits good tracking capabilities under Gaussian process and observation noise while suppressing observation outliers, even in position of leverage. These good performances are obtained only under the validity of the linear approximation of the power system model.
\end{abstract}


\IEEEpeerreviewmaketitle

\section{Introduction}
With the rapid deployment of phasor measurement units (PMU) along with high speed data communication links with large bandwidth, tracking in real-time the dynamics of a power system state is becoming possible. The natural tool to achieve this function is a dynamic state estimator (DSE). Among other benefits, a DSE will provide a situational awareness \cite{Panteli2015} of the system state to the operators and will open the door to real-time wide-area monitoring and control \cite{Taylor2005}. The prerequisites are sufficient observability of the system, robust state estimation and a carefully maintained database. Among the various techniques being proposed in the literature for DSE, the ones based on the Kalman filter have received a great deal of attention. In particular, for nonlinear systems, the extended Kalman filter (EKF) is the most widely used technique. The reader is referred to \cite{NZhou2015} for a comprehensive comparative study. In \cite{EKFKamwa2011}, using the so-called single-machine infinite-bus system and the fourth-order generator model, the authors applied the EKF to estimate the rotor angles, rotor speeds and the direct and quadrature components of the generators internal voltage. In particular, a very important practical problem is addressed where the field voltage is not accessible to metering due to the presence of brushless excitation systems. However, one problem that is rarely addressed in the literature is the robustness of the state estimator to gross measurement errors, which may strongly bias the classical EKF.

In this paper, we develop a new generalized maximum-likelihood (GM)-EKF that exhibits good statistical efficiency under Gaussian  process and observation noise. This characteristic is achieved thanks to the use of the Huber cost function, which is quadratic for small absolute standardized residuals and linear otherwise. Furthermore, our GM-EKF bounds the influence of vertical outliers and bad leverage points, which are identified by means of the projection statistics algorithm \cite{Lmili1996}. The asymptotic covariance matrix of the state estimation error vector is derived via the covariance matrix of the total influence function of the GM-estimator as shown in \cite{Lmili2010}. Simulations carried out on the IEEE 39-bus test system reveal that our robust GM-EKF exhibits good tracking capabilities under Gaussian noise when the linear approximation of the power system model is valid, that is, when the system is not too stressed, resulting in mild nonlinearities.

The paper is organized as follows. Section II introduces the classical problem formulation of the EKF. Section III presents a detailed description of the proposed GM-EKF. Simulation results are discussed in Section IV while some conclusion and future research are provided in Section V.

\section{Classical Problem Formulation}
A discrete-time dynamic system can be modeled as
\begin{equation}
{\boldsymbol{x}_k} = \boldsymbol{f}\left( {{\boldsymbol{x}_{k - 1}},{\boldsymbol{w}_{k - 1}}} \right), \qquad
{\boldsymbol{z}_k} = \boldsymbol{g}\left( {{\boldsymbol{x}_k},{\boldsymbol{v}_k}} \right),
\label{Eq:measurementmodel}
\end{equation}
where $\boldsymbol{x}$\ is the ($n\text{{\fontfamily{cmss}\selectfont x}}1$) state vector; $\boldsymbol{z}$\ is the ($m\text{{\fontfamily{cmss}\selectfont x}}1$) measurements vector; $\boldsymbol{f}(\cdot)$\ and $\boldsymbol{g}(\cdot)$\ are vector-valued nonlinear functions; $\boldsymbol{w}$\ is the ($n\text{{\fontfamily{cmss}\selectfont x}}1$) system process noise vector; $\boldsymbol{v}$\ is the ($m\text{{\fontfamily{cmss}\selectfont x}}1$) measurement noise vector; and no external input is considered. The objective is to estimate the rotor angle $\delta_{i}$ and rotor speed $\omega_{i}$ of the $i$-th synchronous generator, where $i=\{1,...,n_{g}\}$, and $n_{g}$\ is the total number of generators in the system. Thus, $\boldsymbol{x} = {\left[ \omega_{1},...,\omega_{n_{g}},\delta_{1},...,\delta_{n_{g}} \right]^T}$\ and the number of state variables is $n=2n_{g}$. We assume enough measurement redundancy, where the measurement vector is given by
\begin{equation*}
\boldsymbol{z}=[P_{1},...,P_{n_{g}},Q_{1},...,Q_{n_{g}},V_{1},...,V_{n_{b}},\theta_{1},...,\theta_{n_{b}}]^{T}.
\label{eq3}
\end{equation*}
Here $P_{i}$\ and $Q_{i}$\ are real and reactive power injection, respectively; $V_{\ell}$\ and $\theta_{\ell}$ are voltage magnitude and angle, respectively; $\ell=\{1,...,n_{b}\}$ and $n_{b}$\ is the total number of system buses. Thus, $m=2 n_{g}+2n_{b}$. The noise vectors $\boldsymbol{w}_{k}$\ and $\boldsymbol{v}_{k}$ are assumed to be Gaussian, zero-mean and uncorrelated, yielding
\begin{align*}
\mathbb{E}[\boldsymbol{w}_{k}]& = \boldsymbol{0}, & \mathbb{E}[\boldsymbol{w}_{k}\boldsymbol{w}_{j}^{T}]&= \delta_{kj} \boldsymbol{W}_{k}, &  \\
\mathbb{E}[\boldsymbol{v}_{k}]& = \boldsymbol{0}, & \mathbb{E}[\boldsymbol{v}_{k}\boldsymbol{v}_{j}^{T}]&= \delta_{kj} \boldsymbol{R}_{k}, & \mathbb{E}[\boldsymbol{w}_{k}\boldsymbol{v}_{j}^{T}]= \boldsymbol{0},
\end{align*}
where $\mathbb{E}[\cdot]$\ is the expectation operator and $\delta_{kj}$\ is the Kronecker delta. The classical model of a synchronous machine is used, which is expressed as
\begin{align}
\begin{cases}
\dot{\delta}_{i} = (\omega _{i}-\omega _{s}) \\
\dot{\omega}_{i} = \frac{\omega_{s}}{2H_{i}}\big[ P_{m_{i}} - P_{i} - D_{i}(\omega _{i}-\omega _{s}) \big], \\
\end{cases}
\label{Eq:classicalmodel}
\end{align}
\begin{equation*}
P_{i}= G_{ii}E_{i}^{2} + \sum_{j=1,\ j \ne i}^{n_{g}} E_{i}E_{j} \big[ G_{ij}\cos(\delta_{i}-\delta_{j}) + B_{ij}\sin(\delta_{i}-\delta_{j}) \big],
\end{equation*}
where $\delta_{i}$\ is the angular position of rotor in generator $i$\ with respect to generator $1$; $\omega_{i}$\ is the deviation of rotor speed in generator $i$\ relative to that of generator $1$; $\omega_{s}$\ is the synchronous rotor speed; $H_i$\ is the inertia constant; $P_{{m_i}}$\ is the mechanical power input, which is assumed to be constant; $P_{i}$\ is the electrical power injection; $D_i$ is the damping coefficient associated with the $i$-th generator; $E_{i}$\ is the internal voltage magnitude; and $G_{ii}$, $G_{ij}$\ and $B_{ij}$\ are system and load parameters.
\subsection{The standard extended Kalman filter}
The EKF, which is a two step prediction-correction process, can be summarized as follows \cite{Simon2006}. First, the filter is initialized by setting
\begin{align}
\boldsymbol{\hat{x}}_{k|k}^{0} & =\mathop{\mathbb{E}}\big[\boldsymbol{x}_{}^{0}\big], &
\boldsymbol{\Sigma}_{k|k}^{0} & =\mathop{\mathbb{E}}\big[(\boldsymbol{x}-\boldsymbol{\hat{x}}_{k|k}^{0})(\boldsymbol{x}-\boldsymbol{\hat{x}}_{k|k}^{0})^{T}\big].
\end{align}
Then, for $k=\{1,2,...\}$, a \textit{state prediction} followed by a \textit{state correction} are alternatively performed by executing the algorithm described next.

\subsubsection{State prediction}
Calculate the one-step prediction of the system state along with the associated covariance matrix of the state estimation error. Formally, we have
\begin{align}
\boldsymbol{\hat{x}}_{k|k-1} & = \boldsymbol{f}(\boldsymbol{\hat{x}}_{k-1|k-1}), \label{state_predicted} \\
\boldsymbol{\Sigma}_{k|k-1}  & = \boldsymbol{F}_{k-1}^{}\boldsymbol{\Sigma}_{k-1|k-1}^{}\boldsymbol{F}_{k-1}^{T}+\boldsymbol{W}_{k-1}^{}. \label{covariance_predicted}
\end{align}

\subsubsection{State correction} 
Calculate the Kalman filter gain and update the state estimate and the estimation error covariance matrix using
\begin{align}
\boldsymbol{K}_{k} & =\boldsymbol{\Sigma}_{k|k-1}\boldsymbol{H}_{k}^{T}(\boldsymbol{H}_{k}^{}\boldsymbol{\Sigma}_{k|k-1}^{}\boldsymbol{H}_{k}^{T}+\boldsymbol{R}_{k}^{})^{-1}, \label{KalmanGain}\\
\boldsymbol{\hat{x}}_{k|k} & = \boldsymbol{\hat{x}}_{k|k-1}+\boldsymbol{K}_{k}[\boldsymbol{y}_{k} - \boldsymbol{g}(\boldsymbol{\hat{x}}_{k|k-1})], \\
\boldsymbol{\Sigma}_{k|k}^{} & = (\boldsymbol{I}-\boldsymbol{K}_{k}\boldsymbol{H}_{k})\boldsymbol{\Sigma}_{k|k-1}^{}. \label{Eq:errorcovariance}
\end{align}
Notice that $\boldsymbol{F}_{k-1}$\ and $\boldsymbol{H}_{k}$\ are Jacobian matrices defined as
\begin{equation}
\boldsymbol{F}_{k-1} = \frac{\partial \boldsymbol{f}}{\partial \boldsymbol{x}}, \qquad
\boldsymbol{H}_{k} =\frac{\partial \boldsymbol{g}}{\partial \boldsymbol{x}}.
\end{equation}

\section{Proposed robust extended Kalman filter}
The method presented in this section is an extension of the work of Gandhi and Mili \cite{Lmili2010}, where a robust Kalman filter was derived. The GM-EKF is presented next.

\subsection{Derivation of the batch-mode regression form}
Given the initial estimated state vector, $\boldsymbol{\hat x}_{k|k}^{0}$, along with the estimation error covariance matrix, $\boldsymbol{\Sigma}_{k|k}^{0}$, the next time sample is predicted through
\begin{equation}
\boldsymbol{\hat x}_{k|k-1} = \boldsymbol{f} (\boldsymbol{\hat{x}}_{k-1|k-1}),
\label{Eq:prediction}
\end{equation}
\begin{equation}
\boldsymbol{\Sigma}_{k|k-1} = \boldsymbol{F}_{k-1}^{}\boldsymbol{\Sigma}_{k-1|k-1}^{}\boldsymbol{F}_{k-1}^{T} + \boldsymbol{W}_{k-1}^{}.
\label{Eq:predictioncovariance}
\end{equation}

Notice that (\ref{Eq:prediction}) and (\ref{Eq:predictioncovariance}) are exactly the same as (\ref{state_predicted}) and (\ref{covariance_predicted}). Then, we linearize $\boldsymbol{g}\left( {{\boldsymbol{x}_k},{\boldsymbol{v}_k}} \right)$ using a first-order Taylor series expansion around the predicted state vector, $\boldsymbol{\hat{x}}_{k|k-1}$, yielding
\begin{equation}
\boldsymbol{z}_{k} = \boldsymbol{g} (\boldsymbol{\hat{x}}_{k|k-1}) + \boldsymbol{H}_{k} (\boldsymbol{x}_{k} - \boldsymbol{\hat{x}}_{k|k-1}) + \boldsymbol{v}_{k}.
\label{Eq:Taylorexpansion}
\end{equation}

In addition, define
\begin{equation}
\boldsymbol{\hat{x}}_{k|k-1} = \boldsymbol{x}_{k} - \boldsymbol{\delta}_{k|k-1},
\label{eq13}
\end{equation}
where $\boldsymbol{\delta}_{k|k-1}$ is the prediction error and $\mathop{\mathbb{E}} \big[\boldsymbol{\delta}_{k|k-1}^{} \boldsymbol{\delta}_{k|k-1}^{T} \big] = \boldsymbol{\Sigma}_{k|k-1}$. Combining (\ref{Eq:Taylorexpansion}) and (\ref{eq13}), we obtain the batch-mode regression form given by
\small
\begin{equation}
\left[ {\begin{array}{*{20}{c}}
\boldsymbol{z}_{k} - \boldsymbol{g}(\boldsymbol{\hat{x}}_{k|k-1}) + \boldsymbol{H}_{k}\boldsymbol{\hat{x}}_{k|k-1} \\
\boldsymbol{\hat{x}}_{k|k-1}
\end{array}} \right] = \left[ {\begin{array}{*{20}{c}}
\boldsymbol{H}_{k}\\
\boldsymbol{I}
\end{array}} \right]\boldsymbol{x}_{k} + \left[ {\begin{array}{*{20}{c}}
\boldsymbol{v}_{k}\\
-\boldsymbol{\delta}_{k|k-1}
\end{array}} \right],
\label{Eq:batchmodel}
\end{equation}
\normalsize
which can be rewritten in a compact form as
\begin{equation}
\boldsymbol{\tilde{z}}_{k} = \boldsymbol{\tilde{H}}_{k}\boldsymbol{x}_{k} + \boldsymbol{\tilde{e}}_{k}.
\label{Eq:batchmodelcompact}
\end{equation}

The covariance matrix of the error $\boldsymbol{\tilde{e}}_{k}$ is given by 
\begin{equation}
\mathbb{E}\big[ \boldsymbol{\tilde{e}}_{k}\boldsymbol{\tilde{e}}_{k}^{T} \big] = \boldsymbol{\tilde{R}}_{k} =\left[ {\begin{array}{*{20}{c}}
{\boldsymbol{R}_{k}}&\boldsymbol{0}\\
\boldsymbol{0}&\boldsymbol{\Sigma}_{k|k-1}
\end{array}} \right] = \boldsymbol{S}_{k}^{}\boldsymbol{S}_{k}^{T},
\label{Eq:batchmodelcompactcovariance}
\end{equation}
where $\boldsymbol{S}_{k}$ is obtained from the Cholesky decomposition. Note that the dimension of $\boldsymbol{\tilde{z}}_{k}$ is $m'=m+n$.
\subsection{Outlier identification}
If we directly apply the matrix $\boldsymbol{S}_k$ in the prewhitening step, the outliers present in the data will smear over the elements of $\boldsymbol{\tilde{z}}_{k}$ and $\boldsymbol{\tilde{H}}_{k}$. It is thus necessary to detect and suppress the outliers first. This is carried out by using the projection statistics algorithm described in \cite{Lmili1996}. In \cite{Lmili2010}, the projection statistics are calculated using a matrix $\boldsymbol{Z}$ of dimension $(m'\ \text{{\fontfamily{cmss}\selectfont x}}\ 2)$ with the column vectors $\boldsymbol{\tilde{z}}_{k}$\, and $\boldsymbol{\tilde{z}}_{k-1}$. This approach is slightly modified here, due to the presence of the term $\boldsymbol{H}_{k}\boldsymbol{\hat{x}}_{k|k-1}$ in (\ref{Eq:batchmodel}). Consequently, we redefine $\boldsymbol{Z}$ as
\begin{equation}
\boldsymbol{Z}=
\left[ {\begin{array}{*{20}{c cc}}
\boldsymbol{z}_{k-1} - \boldsymbol{g}(\boldsymbol{\hat{x}}_{k-1|k-2})   & \boldsymbol{z}_{k} - \boldsymbol{g}(\boldsymbol{\hat{x}}_{k|k-1}) \\
\boldsymbol{\hat{x}}_{k-1|k-2}                                          & \boldsymbol{\hat{x}}_{k|k-1}
\end{array}} 
\right],
\label{redefinedZ}
\end{equation}
where $\boldsymbol{z}_{k} - \boldsymbol{g}(\boldsymbol{\hat{x}}_{k|k-1})$ is the innovation vector and the dimension of $\boldsymbol{Z}$\ is the same as indicated before. Then, each $i$-th computed projection statistic value, $PS_{i}$, is compared to a given threshold. The flagged outliers are then downweighted using the following weight function: $\varpi_{i} = \min \big(1,\ d^{2}/PS_{i}^{2}\big)$, where we pick $d=1.5$\ to yield good statistical efficiency at the Gaussian distribution without increasing too much the bias of the GM-estimator under contamination. 

\subsection{Robust prewhitening}
The prewhitening is then performed by pre-multiplying (\ref{Eq:batchmodelcompact}) by $\boldsymbol{S}_{k}^{-1}$, resulting in
\begin{equation}
\boldsymbol{S}_{k}^{-1} \boldsymbol{\tilde{z}}_{k} = \boldsymbol{S}_{k}^{-1} \boldsymbol{\tilde{H}}_{k} \boldsymbol{x}_{k} + \boldsymbol{S}_{k}^{-1} \boldsymbol{\tilde{e}}_{k}.
\label{Eq:prewhiteningbatchmodel}
\end{equation}

Note from (\ref{Eq:batchmodelcompactcovariance}) that $\boldsymbol{S}_{k}^{}$ is robust because $\boldsymbol{\Sigma}_{k|k-1}$ is robust and $\boldsymbol{R}_{k}$ is just the given covariance matrix of the measurements noise. Equation (\ref{Eq:prewhiteningbatchmodel}) can be further organized in a compact form as
\begin{equation}
\boldsymbol{y}_{k} = \boldsymbol{A}_{k} \boldsymbol{x}_{k} + \boldsymbol{\xi}_{k}.
\label{Eq:finalbatchmodel}
\end{equation}

\subsection{Robust filtering and solution}
The generalized maximum-likelihood estimator that bounds the influence of outliers on the state estimates minimizes an objective function given by
\begin{equation}
J(\boldsymbol{x}) = \sum_{i=1}^{m} {\varpi_{i}^{2}}\ \rho(r_{S_{i}}),
\label{Eq:objectivefunction}
\end{equation}
where $\rho(\cdot)$\ is a nonlinear function of the residuals. The Huber $\rho$-function is used here. It is defined as
\begin{equation}
\rho(r_{S_{i}}) = 
\begin{cases}
0.5\ r_{S_{i}}^{2}, & \qquad \big|r_{S_{i}}\big|<c \\
c\ \big|r_{S_{i}}\big|-0.5\ c^{2}, & \qquad \text{otherwise}.
\end{cases}
\label{Eq:huberfunction}
\end{equation}
Here we set $c=1.5$ to obtain high statistical efficiency under Gaussian noise. The standardized residuals, $r_{S_{i}}$, are defined as $r_{S_{i}} = r_{i}/(s\ \varpi_{i})$, where $s$\ is a robust estimator of scale expressed as $s=1.4826\cdot b_{m'}\cdot \mathrm{median}_{i}\big|r_{i}\big|$; the residuals are given by $r_{i} = y_{i} - \boldsymbol{a}_{i}^{T} \boldsymbol{\hat{x}}$, and $\boldsymbol{a}_{i}$ is the $i$-th column vector of the matrix $\boldsymbol{A}_{k}^{T}$. Note that $\varpi_{i}$ is used to bound the influence of position while the Huber $\rho$-function is chosen to bound the influence of the residuals. To minimize $J(\boldsymbol{x})$ given by (\ref{Eq:objectivefunction}), one takes its partial derivative and sets it equal to zero, yielding
\begin{equation}
\frac {\partial J(\boldsymbol{x})} {\partial \boldsymbol{x}} = \sum_{i=1}^{m} -\frac {\varpi_{i} \boldsymbol{a}_{i}} {s} \psi(r_{S_{i}}) = \boldsymbol{0},
\label{Eq:objectivefunctionpartialderivative}
\end{equation}
where $\psi(r_{S_{i}}) = \partial \rho(r_{S_{i}})/r_{S_{i}}$. The state vector correction is then computed via the iteratively reweighted least squares (IRLS) algorithm, which is expressed as
\begin{equation}
\boldsymbol{\hat{x}}_{k|k}^{(j + 1)} = \left( \boldsymbol{A}_{k}^{T} \boldsymbol{Q}_{}^{(j)} \boldsymbol{A}_{k}^{} \right)^{-1} \boldsymbol{A}_{k}^{T} \boldsymbol{Q}_{}^{(j)} \boldsymbol{y}_{k},
\label{Eq:IRLS}
\end{equation}
where $\boldsymbol{Q} = \mathrm{diag}\{ q(r_{S_{i}}) \}$ and $q(r_{S_{i}}) = \psi(r_{S_{i}})/r_{S_{i}}$. The algorithm converges when $||\boldsymbol{\hat{x}}_{k|k}^{(j + 1)}-\boldsymbol{\hat{x}}_{k|k}^{(j)}||_{\infty} \le$\ IRLS tolerance.

\subsection{Covariance calculation and updating}
Upon convergence of the IRLS algorithm, the estimation error covariance matrix, $\boldsymbol{\Sigma}_{k|k}$, needs to be updated, thus allowing the state forecasting to be performed for the next time sample. Unlike the EKF, which uses (\ref{Eq:errorcovariance}) to compute $\boldsymbol{\Sigma}_{k|k}$, the GM-EKF uses an expression derived from the total influence function \cite{Lmili2010}, which is given by
\begin{equation}
\boldsymbol{\Sigma}_{k|k} = \frac {\mathbb{E}_{\Phi} \big[ \psi^{2}(r_{S_{i}}) \big]} { \big\{ \mathbb{E}_{\Phi} \big[ \psi^{\prime}(r_{S_{i}}) \big] \big\}^{2}} (\boldsymbol{A}_{k}^{T} \boldsymbol{A}_{k}^{})^{-1} (\boldsymbol{A}_{k}^{T} \boldsymbol{Q}_{\varpi}^{} \boldsymbol{A}_{k}^{})^{} (\boldsymbol{A}_{k}^{T} \boldsymbol{A}_{k}^{})^{-1}
\label{Eq:updatingSigmafinal},
\end{equation}
where $\boldsymbol{Q}_{\varpi} = \mathrm{diag}(\varpi_{i}^{2})$. When $c=1.5$, the value of the first term in the right-hand side of (\ref{Eq:updatingSigmafinal}) is equal to 1.0369.

\subsection{Tuning the GM-EKF}
\begin{figure}[htb!]
\centering
\scriptsize
\begin{tikzpicture}[node distance=1.2cm]
\node (start) [startstop] {\textbf{Init.:} $\boldsymbol{\hat{x}}_{k|k}^{0}$, $\boldsymbol{\Sigma}_{k|k}^{0}$, $\boldsymbol{W}_{k-1}$\ and $\boldsymbol{R}_{k}$};
\node (pro0)  [process, below of=start] {$\boldsymbol{\hat{x}}_{k|k-1}$, $\boldsymbol{\Sigma}_{k|k-1}$, $\boldsymbol{g}(\boldsymbol{\hat{x}}_{k|k-1})$, and $\boldsymbol{H}_{k}$ };
\node (pro1)  [process,   below of=pro0,  yshift=0.0cm] {(step 1) $\boldsymbol{\tilde{z}}_{k}$ and $\boldsymbol{\tilde{H}}_{k}$};
\node (pro2)  [process,   below of=pro1,  yshift=0.0cm] {(step 2) $\boldsymbol{PS}$, $\boldsymbol{\varpi}$ and $\boldsymbol{Q}_{\varpi}$ };
\node (pro3)  [process,   below of=pro2,  yshift=0.0cm] {(step 3) $\boldsymbol{\tilde{R}}$, $\boldsymbol{S}_{k}$, $\boldsymbol{y}_{k}$ and $\boldsymbol{A}_{k}$ };
\node (pro4)  [startstop, below of=pro3,  yshift=0.0cm] {\textbf{Init. IRLS:}\ $\boldsymbol{\hat{x}}_{k|k}^{(j)}$, $j=0$ };
\node (pro5)  [process,   below of=pro4,  yshift=0.0cm] {(step 4) $\boldsymbol{r}_{S}$, $\boldsymbol{Q}$ and $\boldsymbol{\hat{x}}_{k|k}^{(j+1)}$ };
\node (dec1)  [decision,  below of=pro5,  yshift=-1.5cm] {$\norm{\boldsymbol{\hat{x}}_{k|k}^{(j+1)}-\boldsymbol{\hat{x}}_{k|k}^{(j)}}<\text{tol.}$};
\node (pro6a) [process,   left  of=dec1,  xshift=-2.2cm] {$\boldsymbol{\hat{x}}_{k|k}\leftarrow \boldsymbol{\hat{x}}_{k|k}^{(j+1)}$};
\node (pro6b) [process,   right of=dec1,  xshift=+1.9cm] {$j=j+1$};
\node (pro7)  [process,   above of=pro6a, yshift=+0.8cm] {(step 5) $\boldsymbol{\Sigma}_{k|k}$};
\draw [arrow] (start) -- (pro0);
\draw [arrow] (pro0) -- (pro1);
\draw [arrow] (pro1) -- (pro2);
\draw [arrow] (pro2) -- (pro3);
\draw [arrow] (pro3) -- (pro4);
\draw [arrow] (pro4) -- (pro5);
\draw [arrow] (pro5) -- (dec1);
\draw [arrow] (dec1) -- (pro6a);
\draw [arrow] (dec1) -- (pro6b);
\draw [arrow] (dec1) -- node[anchor=south] {Y} (pro6a);
\draw [arrow] (dec1) -- node[anchor=south] {N} (pro6b);
\draw [arrow] (pro6b) |- (pro5);
\draw [arrow] (pro6a) -- (pro7);
\draw [arrow] (pro7) |- (pro0);
\tikzstyle{process} = [rectangle, minimum width=3cm, minimum height=1cm, text centered, text width=3cm, draw=black, fill=orange!30]
\end{tikzpicture}
\caption{GM-EKF algorithm flowchart.}
\label{flowchart}
\end{figure}
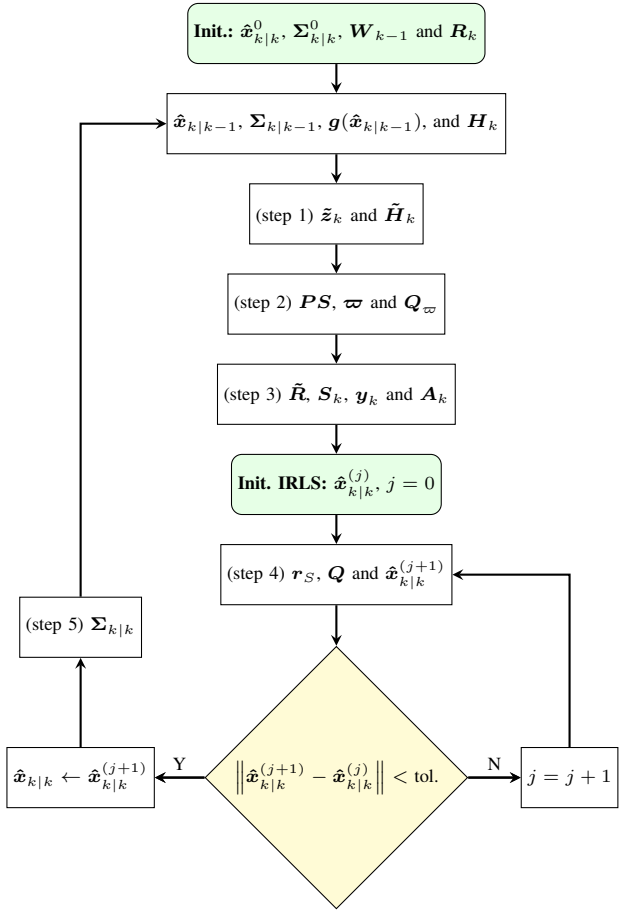

To tune the filter, three parameters can be adjusted, namely $c$, $d$\ and the IRLS tolerance. The latter should not be set too small. According to our experience, $0.01$\ is a good choice; decreasing this tolerance does not bring relevant numerical improvement, but increases the computation time. The parameter $c$\ is the breakpoint of the Huber $\rho$-function, whose value determines the trade-off that we wish to have between a least-squares and a least-absolute-value fit. Indeed, if $c\rightarrow 0$, the Huber $\rho$-function tends to the least-absolute-value $\rho$-function and if $c\rightarrow \infty$, it tends to the least-squares $\rho$-function. As for the parameter $d$ of the weight function, it determines the statistical efficiency of the projection statistics method at the Gaussian distribution and the robustness of the GM-estimator. Decreasing this parameter shrinks the dimensions of the ellipse that determines the confidence area in the two-dimensional case. As a result, good measurements are unduly downweighted, which yields a decrease in the statistical efficiency of the method. On the other hand, increasing $d$ will decrease the breakdown point of the GM-estimator. As for the parameter $b_{m'}$\, Croux and Rousseeuw \cite{Croux1992} suggest to set it equal to the values displayed in Table \ref{tab22} if $m' \le 9$, and to $b_{m'}=[m'/(m'-0.8)]$ otherwise. Finally, Fig. \ref{flowchart} presents a flowchart of the GM-EKF algorithm.
\begin{table}[htb!]
\caption{Parameter $b_{m'}$\ for $m'\le 9$.}
\centering
\setlength{\tabcolsep}{0.5em}
\begin{tabular}{c c c c c c c c c}
\hline \hline
$m'$ & $2$ & $3$ & $4$ & $5$ & $6$ & $7$ & $8$ & $9$ \\
\hline
$b_{m'}$ & $1.196$ & $1.495$ & $1.363$ & $1.206$ & $1.200$ & $1.140$ & $1.129$ & $1.107$ \\
\hline
\end{tabular}
\label{tab22}
\end{table}

\section{Simulation results}
\subsection{Comparison between EKF, GM-EKF and UKF under ideal conditions}
\begin{figure}[htb!]
\centering
\fbox{\includegraphics[width=8.5cm]{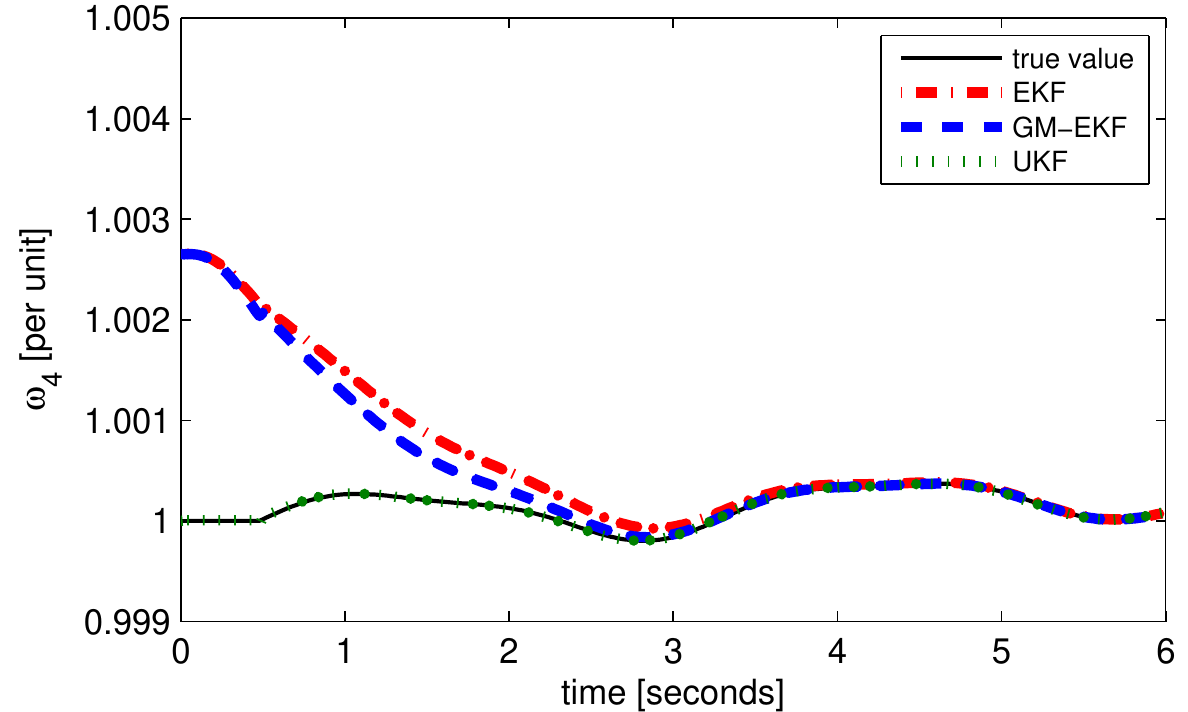}}
\caption{Rotor speed of Generator 4 after the applied disturbance at $t=0.5s$.}
\label{ex1-w4}
\end{figure}

In the simulations, we assume that $\boldsymbol{W}_{k-1}=\boldsymbol{R}_{k}=\mathrm{diag}\{10^{-4}\}$, which does not represent a stringent condition. The presence of outliers is not considered at this point. For comparison purpose, we also simulate the unscented Kalman filter (UKF), although no theoretical background has been previously presented; the reader is referred to \cite{Uhlmann2004,Terzija2011,UKFKamwa2011} for details. The rotor speed versus time is displayed in Fig. \ref{ex1-w4}. We pick the plot of Generator 4 as an example and omit the remaining plots since they do not bring any additional qualitative information. Clearly, it can be seen from Fig. \ref{ex1-w4} that all the three methods are able to track well system dynamics.

\begin{figure}[htb!]
\centering
\fbox{\includegraphics[width=8.5cm]{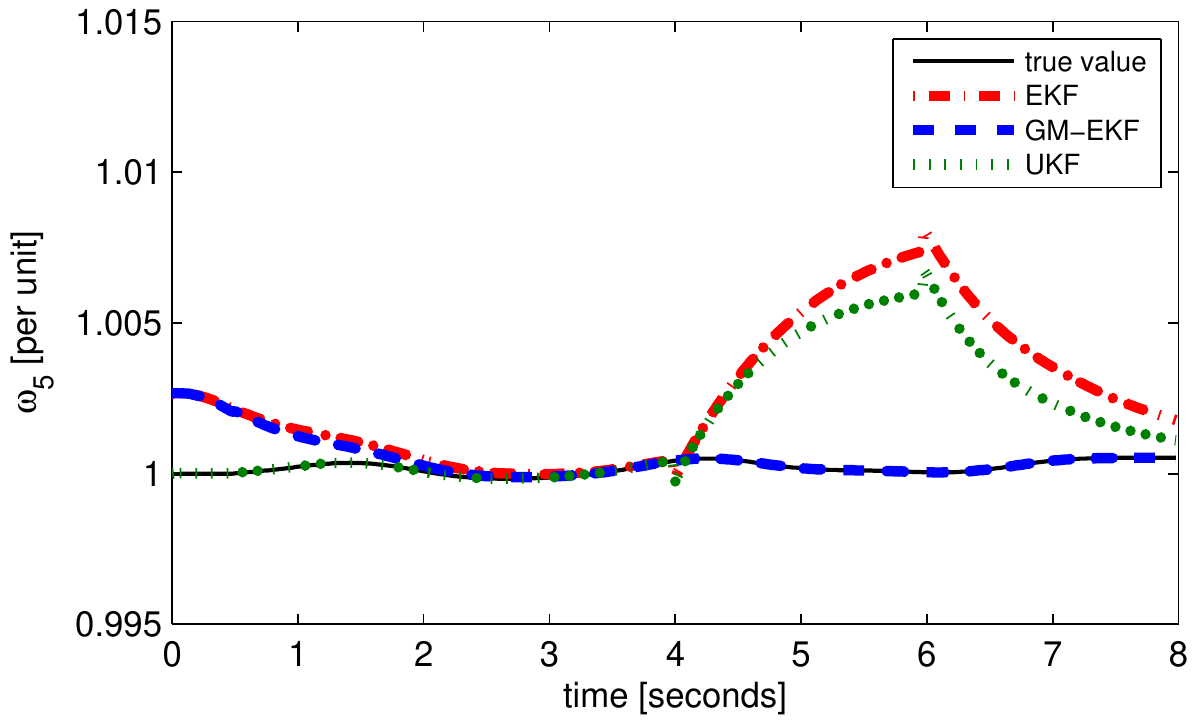}}
\caption{Rotor speed of Generator 5 with loss of PMU \#5 from $4$ to $6s$.}
\label{ex2-w5}
\end{figure}

\begin{figure}[htb!]
\centering
\fbox{\includegraphics[width=8.5cm]{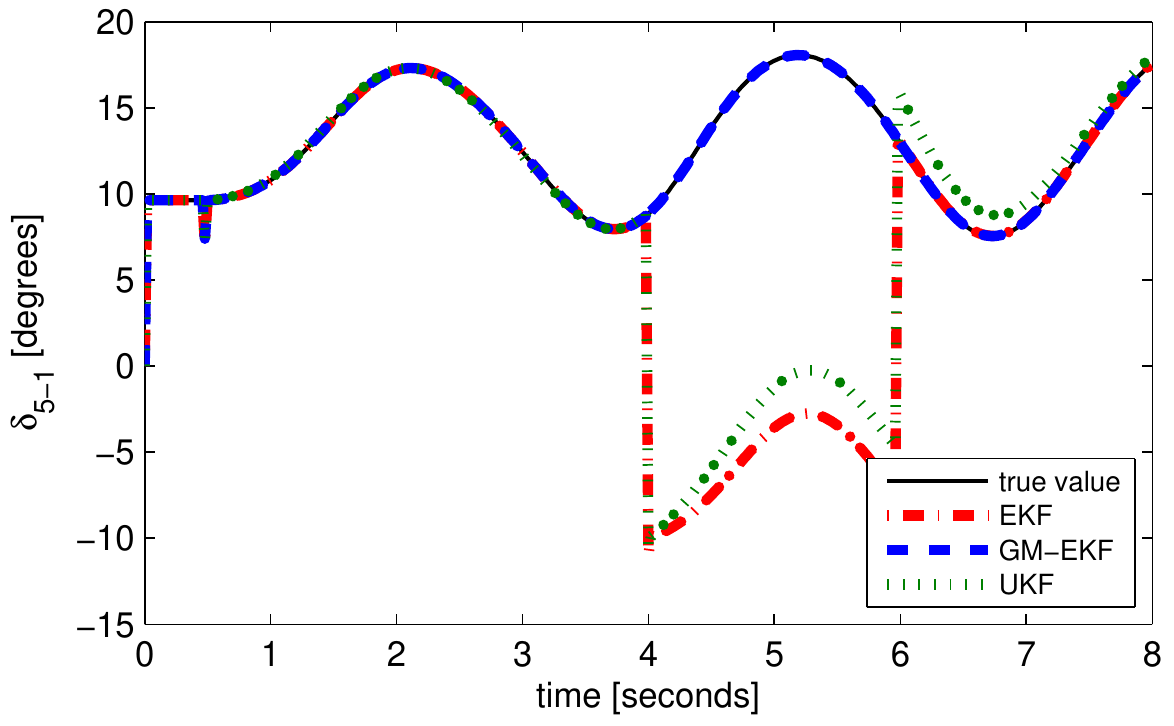}}
\caption{Rotor angle of Generator 5 with loss of PMU \#5 from $4$ to $6s$.}
\label{ex2-delta5}
\end{figure}

\subsection{Momentary loss of communication link}
The communication link with the PMU placed at Bus 34 where Generator 5 is connected is supposed to be lost from $4$\ to $6s$. Therefore, the measurement set $\{P_{5}, Q_{5}, V_{5}, \theta_{5}\}$ becomes unavailable during this time frame; here, their values are set equal to zero for simulation purpose. Although such an event has a low probability of occurrence, it is of interest to investigate its effects on the estimation results. Fig. \ref{ex2-w5} and \ref{ex2-delta5} illustrate the estimation of the state variables of Generator 5. As we can observe, the non-robust methods, namely the EKF and the UKF, are not capable of handling such condition. Note that although we are not showing the remaining plots, the estimation related to other generators are also strongly biased by the outliers. By contrast, the proposed GM-EKF downweights these outliers to the point to nearly suppress their effect on the state estimates.
\subsection{Measurements with bad data}
\begin{figure}[htb!]
\centering
\fbox{\includegraphics[width=8.5cm]{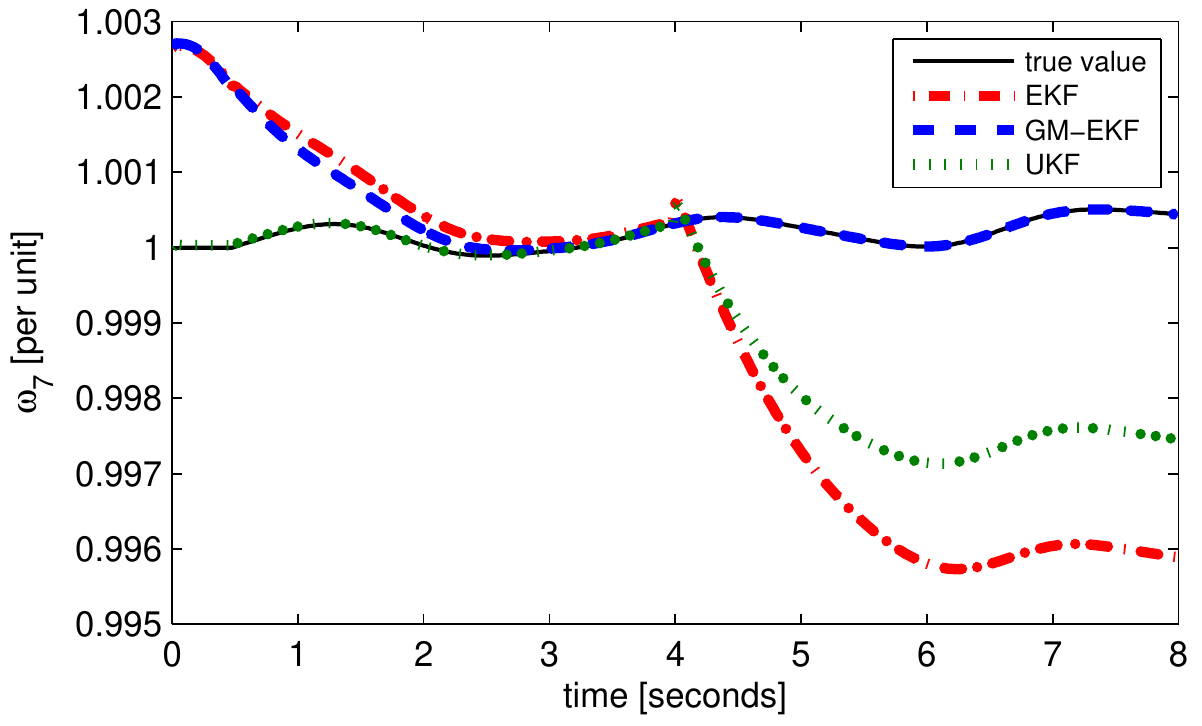}}
\caption{Rotor speed of Generator 7 with an outlier placed on $Q_{7}$ after $t=4s$.}
\label{ex3-w7}
\end{figure}

\begin{figure}[htb!]
\centering
\fbox{\includegraphics[width=8.5cm]{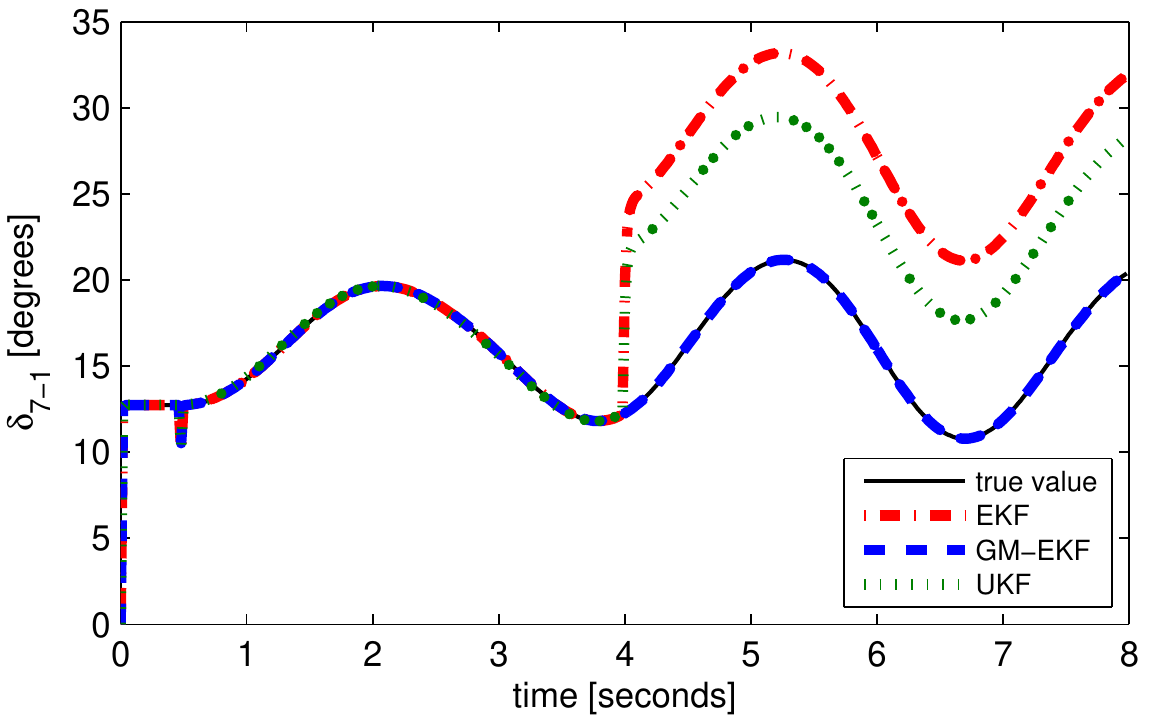}}
\caption{Rotor angle of Generator 7 with an outlier placed on $Q_{7}$ after $t=4s$.}
\label{ex3-delta7}
\end{figure}

In this case, we set $Q_{7}=10$\ at $t=4s$, which represents a gross measurement error since the reactive power injection is below $2.2$ per-unit for all the generators. The simulation results are displayed in Figs. \ref{ex3-w7} and \ref{ex3-delta7}. We observe that the larger the magnitude of the outlier is, the worst the estimation results are when using EKF and UKF. As for the GM-EKF, the outlier is strongly downweighted, resulting in a good state estimate. The overall estimation error for the three simulated cases is presented in Table \ref{tab_overallestimationerrors}. For comparison, we present in Table \ref{tab_computationtime} the computing times for all the simulated cases although the Matlab code has not been optimized. Here, we have run the algorithm 200 times for each case and calculated the average times.

\begin{table}[htb!]
\caption{Overall estimation errors.}
\centering
\setlength{\tabcolsep}{1.5em}
\begin{tabular}{c| c c c}
\hline \hline
$\ $Method$\ $ & EKF & GM-EKF & UKF \\
\hline
Case 1 & $0.0169$ & $0.0141$ & $0.0009$  \\
Case 2 & $0.0235$ & $0.0106$ & $0.0101$  \\
Case 2 & $0.0275$ & $0.0106$ & $0.0098$  \\
\hline
\end{tabular}
\label{tab_overallestimationerrors}
\end{table}

\begin{table}[htb!]
\caption{Computational time.}
\centering
\setlength{\tabcolsep}{1.5em}
\begin{tabular}{c| c c c}
\hline \hline
$\ $Method$\ $ & EKF & GM-EKF & UKF \\
\hline
Case 1 & $1.3482s$ & $2.0398s$ & $0.9266s$  \\
Case 2 & $1.7859s$ & $2.6811s$ & $1.2986s$  \\
Case 2 & $1.6956s$ & $2.6662s$ & $1.2328s$  \\
\hline
\end{tabular}
\label{tab_computationtime}
\end{table}

\section{Conclusions}
A robust GM-EKF has been derived and its performance has been assessed on the IEEE 39-bus system.  It has been shown that the developed filter is able to track power system dynamics when system nonlinearities are not too strong.  Another interesting property of the GM-EKF is that it exhibits high statistical efficiency under Gaussian noise while suppressing gross errors in PMU measurements, be they vertical outliers or bad leverage points.  However, our filter has several weaknesses. Firstly, it is unable to cope with innovation outliers due to smearing effect. Secondly, it loses its tracking ability under strong system nonlinearities, which occur when the power system is stressed or under large variation of the system state following a severe disturbance.  Thirdly, it presents poor performance under non-Gaussian noise.  This is a major drawback since the Gaussian assumption has been invalidated by a recent PNNL report \cite{Huang2015b}. Analyzing actual field PMU data, that report shows that the measurement errors of the voltage and current magnitudes follow bimodal Gaussian-mixture distributions. For these distributions, other types of robust filters should be developed. This is the topic of future research.

\section*{Acknowledgment}
This work was supported by CAPES Foundation - Ministry of Education of Brazil under grant BEX13594/13-3, and by National Natural Science Foundation of China under grants 61170016 and 61373047.

\ifCLASSOPTIONcaptionsoff
  \newpage
\fi

\end{document}